\documentstyle[aps,preprint,epsfig]{revtex}

\begin{document}

  \title{Temperature behavior of vortices of a 3D thermoconducting viscous 
   fluid}
  \author{V. Grassi, R.A. Leo, G. Soliani and P. Tempesta\\
   Dipartimento di Fisica dell'Universit\`a, 73100 Lecce, Italy,\\
   and Istituto Nazionale di Fisica Nucleare, Sezione di Lecce, Italy}
\maketitle

\begin{abstract}
The Navier-Stokes-Fourier model for a 3D thermoconducting viscous fluid, where 
the evolution equation for the temperature $T$ contains a term proportional to 
the rate of energy dissipation, is investigated analitically at the light of 
the rotational invariance property. Two cases are considered: the Couette flow 
and a flow with a radial velocity between two rotating impermeable and porous 
coaxial cylinders, respectively. In both cases, we show the existence of a 
maximum value of $T$, $T_{\rm{max}}$, when the difference of temperature 
$\Delta T=T_2-T_1$ on the surfaces of the cylinders is assigned. The role of 
$T_{\rm{max}}$ is discussed in the context of different physical situations.
\end{abstract}
PACS numbers: 47.32.Cc
\vspace{1cm}

In Ref. \cite{G}, a thermoconducting incompressible viscous fluid system, named 
Navier-Stokes-Fourier (NSF) model, is presented. In 3D, this model is governed 
by the equations
\begin{eqnarray}
&&{\bf{u}}_t+{\bf{u}}\cdot\nabla{\bf{u}}+\nabla({p\over\rho})-
  \nu\nabla^2{\bf{u}}=0,\label{1}\\
&&\nabla\cdot{\bf{u}}=0,\label{2a}\\
&&T_t+{\bf{u}}\cdot\nabla T-k_H\nabla^2 T=
  {\eta\over{\rho C_p}}{\sum_{i,j}}(\partial_i u_j+\partial_j u_i)^2,\label{2}
\end{eqnarray}
where ${\bf{u}}$ is the velocity field, $T=T(x,y,z,t)$ is a passive scalar 
identified by the fluid temperature \cite{Beh}, $p=p(x,y,z,t)$ the fluid 
pressure, $\rho$ the fluid density, $\eta$ the dynamic viscosity, 
$\nu=\eta/\rho$ the kinematic viscosity, $C_p$ the heat capacity, $\kappa$ the 
heat conductivity, $k_H=\kappa/(\rho C_p)$ the thermal diffusivity, and 
$\partial_i\equiv\partial/\partial x_i$ ($x_1\equiv x$, $x_2\equiv y$, 
$x_3\equiv z$).\par
The r.h.s. in (\ref{2}) is related to the rate of energy dissipation 
$\varepsilon=2\nu S^2$, where $S^2=S_{ij}S_{ij}$ and 
$S_{ij}=(1/2)(\partial_i u_j+\partial_j u_i)$ are the strain matrix 
elements. Among the many questions arising in the study of the model 
(\ref{1})-(\ref{2}), such as for instance the investigation of statistical
solutions and the onset of turbulence \cite{C}, the presence of a dissipation
term in the equation for the temperature deserves a special attention. At the
best of our knowledge, in three spatial dimensions Eqs. (\ref{1})-(\ref{2})
have never been studied by exploiting an analytic procedure. In this Letter
we discuss some new thermal effects described by exact solutions of the vortex
type of the NSF model. In particular, we have analyzed a steady Taylor-Couette 
flow for an incompressible fluid entrapped between two rotating coaxial 
impermeable cylinders, and a flow with an additional radial component of the 
velocity, when the walls of the cylinders are porous (for technical details on
the experimental devices, see \cite{ML,LDM}). As is well-known, the 
hydrodynamic problems connected with the system of two coaxial rotating 
cylinders has been intensively studied from an experimental and theoretical 
point of view since the work of Couette (1890) \cite{Cou}. The main motivation 
was the discovery by Taylor of unstable flow regimes consisting in toroidal 
vortices (1923), which appear for high values of the Taylor number \cite{S}. 
This number can be defined in several forms. Following \cite{LDM} we will use 
$Ta=r_1\Omega_1 d/\nu$, where $r_1$ is the radius of the inner cylinder, $d$ 
the width of the gap between the two cylinders, $\Omega_1$ the rotational 
velocity of the inner cylinder. In our context, we will consider low Taylor 
numbers ($Ta\sim 100\div 200$), in order to preserve the geometrical 
symmetries of the model.\par
In the two cases of impermeable and porous cylinders, we show the existence of 
a value $T_{\rm{max}}$ for the temperature field associated with the fluid, 
when the difference of temperature $\Delta T=T_2-T_1$ on the surfaces of the 
cylinders is assigned for asymptotic values of time. This maximum appears as a 
consequence of a mechanism of energy dissipation in heat, due to viscous 
effects, and depends on the width of the gap between the cylinders, on their 
relative angular velocity, and on the fixed value of $\Delta T$. Remarkably, 
we found that the position of $T_{\rm{max}}$ as a function of the radial 
coordinate $r$ can vary continuously in the gap $r_1\leq r\leq r_2$ between 
the two cylinders by varying the parameter $\Delta T$.\par
In the NSF model, $\rho$ is taken to be constant. Although this assumption 
could appear as a too severe restriction on the possible modelling of realistic
situations, nevertheless the NSF model deserves to be analyzed as a 
guide-framework for the investigation of less ideal systems, such as the 
Rayleigh and the Lorentz models, which describe incompressible fluid motions 
where thermal phenomena are more significant \cite{G}.\par
We found that the NSF model allows Lie-point symmetries generated by 
infinitesimal operators giving finite group transformations and exact 
solutions via the corresponding reduced equations. (The technical machinery 
yielding these reductions is outlined in \cite{GLST1,GLST2}). Many of these 
symmetries have a nice geometric interpretation: they express the invariance 
of the NSF system under rotations, translations, Galilean boosts and scale 
transformations. Therefore, as one expects, their expressions do not 
explicitly contain the viscosity $\nu$. Nevertheless, such symmetries hold 
only in the viscous case, in the sense that the corresponding symmetries for 
the inviscid situation cannot be reproduced by those for $\nu\neq 0$ by 
setting $\nu=0$.\par
Here we shall limit ourselves to deal with the rotational symmetry defined by 
the generator
\begin{equation}
V_R=y\partial_x-x\partial_y+u_2\partial_{u_1}-u_1\partial_{u_2}.
\label{3}
\end{equation}
A more systematic treatment of the symmetry properties of the NSF model together
with the investigation of the role of the boundary conditions will be reported
in a separated paper. The operator (\ref{3}) gives rise to a reduced NSF (RNSF) 
system whose last equation is
\begin{eqnarray}
&&\theta_t+U_1\theta_r+U_3\theta_z-{\eta\over{\rho C_p}}\left(2U_{1r}^2+
  2{{U_1^2}\over{r^2}}+U_{1z}^2+U_{2r}^2\right.-\nonumber\\
&&\left.2{{U_2 U_{2r}}\over r}+{{U_2^2}\over{r^2}}+U_{2z}^2+U_{3r}^2+2U_{3z}^2+
  2U_{1z}U_{3r}\right)-\nonumber\\
&&k_H\left(\theta_{rr}+{{\theta_r}\over r}+\theta_{zz}\right)=0,\label{r5}
\end{eqnarray}
where $z$, $t$, $r=\sqrt{x^2+y^2}$ are independent symmetry variables and 
$U_1$, $U_2$, $U_3$, $\Pi$, $\theta$, related to the original variables by
\begin{eqnarray}
&&U_1=u_1\cos\varphi+u_2\sin\varphi,\;
  U_2=-u_1\sin\varphi+u_2\cos\varphi,\nonumber\\
&&U_3=u_3,p=\Pi,T=\theta,
\label{4}
\end{eqnarray}
are dependent variables expressed in terms of $z$, $t$, $r$. Here 
$\cos\varphi=x/r$, $\sin\varphi=y/r$, $U_1$, $U_2$ can be interpreted as the 
radial and the azimuthal components of the (reduced) velocity {\bf{U}}, and 
$U_3$ is the component along the $z$-axis.\par
We find that the RNSF system admits two solutions which are, respctively, 
{\it{i)}} the circular Couette flow (in which only the azimuthal component 
$U_2$ of {\bf{U}} is different from zero), and {\it{ii)}} a flow where all the 
three components of {\bf{U}} are different from zero. In this case $U_3$ turns 
out to be time dependent in such a way that $U_3\rightarrow 0$ exponentially 
when the adimensional variable $\mu_0 t$ goes to infinity (see (\ref{19a})). 
In the original velocity components $u_1$, $u_2$ and $u_3$ the circular 
character of the Couette flow appears, while the flow of case {\it{ii)}} shows 
a spiral behavior.\par
{\it{Case i)}}\par
A simple but notable solution to the RNSF system is obtained by choosing 
$U_1=U_3=0$ and
\begin{equation}
U_2={\alpha\over r}+\beta r,
\label{10}
\end{equation}
where $\alpha,\beta$ are constants. Then
\begin{equation}
p=-{{\alpha^2\rho}\over{2r^2}}+{1\over 2}\beta^2\rho r^2+
2\alpha\beta\rho\ln r,
\label{11}
\end{equation}
apart from an arbitrary function of time, while the temperature $T$ is 
provided by
\begin{eqnarray}
&&T=\exp(-\lambda_0 t)\times\nonumber\\
&&\left[c_1 J_0\left(\sqrt{{{\lambda_0-\lambda_1}\over{k_H}}}r\right)+
  c_2 Y_0\left(\sqrt{{{\lambda_0-\lambda_1}\over{k_H}}}r\right)\right]
  \times\nonumber\\
&&\left[c_3\cos\sqrt{{{\lambda_1}\over{k_H}}}z+
  c_4\sin\sqrt{{{\lambda_1}\over{k_H}}}z\right]+T_d^I,
\label{14}
\end{eqnarray}
where 
\begin{equation}
T_d^I=\lambda_2\ln r-{{\alpha^2\eta}\over{k_H\rho C_p}}{1\over{r^2}}+T_0^I
\label{15}
\end{equation}
is the contribution to the temperature coming from the energy dissipation
present in (\ref{2}) and $\lambda_0>\lambda_1>0$, $\lambda_2$, $T_0^I$ and  
$c_j$ are constants. Here $J_0$ and $Y_0$ are the Bessel functions of the 
first and the second kind, respectively (see \cite{AS}, p. 358).\par
The original velocity components corresponding to the reduced velocity 
${\bf{U}}\equiv (0,{\alpha\over r}+\beta r,0)$ are
\begin{equation}
u_1=-\left({\alpha\over{r^2}}+\beta\right)y,\;
u_2=\left({\alpha\over{r^2}}+\beta\right)x,\;
u_3=0,
\label{16}
\end{equation}
from which we have $r=r_0=const$. Equations (\ref{16}) are solved by 
$x=r_0\cos Kt$, $y=r_0\sin Kt$, $z=const$ where 
$K=\alpha/r_0^2+\beta$.\par
The reduced solution (\ref{10}) is pertinent to a steady flow where the fluid 
occupies the gap $r_1\leq r\leq r_2$ between two coaxial cylinders of radii 
$r_1$ and $r_2$ rotating with (constant) angular velocities $\Omega_1$ and 
$\Omega_2$ \cite{A}. Under the hypothesis of no-slip condition on the 
cylinders, $\alpha$ and $\beta$ take the form 
$\alpha=[r_1^2 r_2^2(\Omega_2-\Omega_1)]/(r_1^2-r_2^2)$,
$\beta=(r_1^2\Omega_1-r_2^2\Omega_2)/(r_1^2-r_2^2)$, while the vorticity 
$\mbox{\boldmath{$\omega_R$}}$ associated with the (reduced) velocity 
(\ref{10}) coincides with the vorticity $\mbox{\boldmath{$\omega$}}$ for the 
velocity (\ref{16}).\par
\begin{figure}[htb]
\begin{center}
\epsfig{file=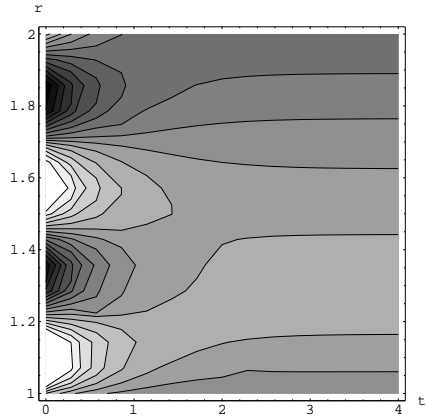,height=5.6cm,width=5.6cm}
\caption{Temperature field for the motion of a fluid between two coaxial
         rigid and impermeable cylinders with radii $r_1=1$ and $r_2=2$ for 
	 $z=0.5$ ($\alpha=1$).}\label{fig1}
\end{center}
\end{figure}
\begin{figure}[htb]
\begin{center}
\epsfig{file=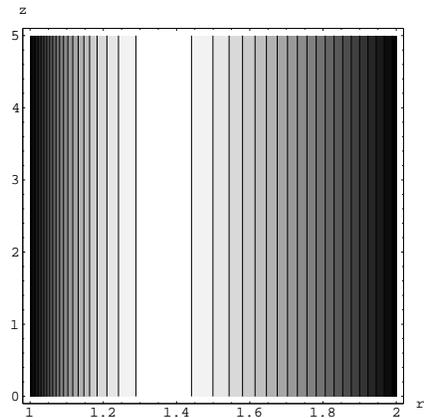,height=5.6cm,width=5.6cm}
\caption{Asymptotic behavior of the thermal field for $1\leq r\leq 2$, 
	 $0\leq z\leq 5$, when $\Delta T=0$ ($Ta=150$).}\label{fig1a}
\end{center}
\end{figure}

In Fig. \ref{fig1} we show the behavior of the temperature field, as being
described by the solution (\ref{14}), as a function of $t$ and $r$ (brighter
regions correspond to higher values of $T$). By exploiting the scale 
invariance of the NSF system we can express the physical constants in an 
adimensional form \cite{B}. Hereafter we will assume $\nu=10^{-2}$, $\rho=1$, 
$\eta=10^{-2}$, $C_p=1$, $\kappa=6\times 10^{-3}$. After a transient phase, 
only the contribution (\ref{15}), due to the dissipative term in (\ref{2}) 
survives. In this situation we have a maximum of $T$ for any $t$ in the gap 
between the two cylinders. This maximum, as shown in Fig. \ref{fig1a} for the 
asymptotic case, can continuously vary by varying $\Delta T$ (or equivalently 
$\lambda_2$).\par
%-------------------------------------------------------------------------------
%-------------------------------------------------------------------------------
\begin{figure}[htb]
\begin{center}
\epsfig{file=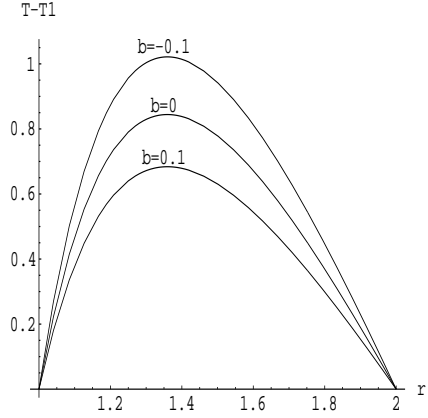,height=5.6cm,width=5.6cm}
\caption{Behavior of the thermal field as a function of the radial distance 
         between the two rigid and impermeable cylinders for some values of 
	 the parameter $b={{\Omega_2}\over{\Omega_1}}$ with $r_1=1$ and 
	 $r_2=2$ ($Ta=150$).}\label{fig1bis}
\end{center}
\end{figure}

%-------------------------------------------------------------------------------
In Fig. \ref{fig1bis}, we plot $T-T_1$ vs. $r$ for different values of 
$b={{\Omega_2}\over{\Omega_1}}$, in the case $\Delta T=0$. We point out that
when the cylinders rotate in the opposite sense ($b<0$), $T_{\rm{max}}$ 
increases. This is related to a growth of the energy dissipation induced by 
friction effects.\par
%-------------------------------------------------------------------------------
%-------------------------------------------------------------------------------
{\it{Case ii)}}\par
Another interesting vortex solution of the NSF equations (\ref{1})-(\ref{2}) 
can be found by putting
\begin{equation}
U_1={{\gamma\nu}\over r},\;U_2=Ar^{\gamma+1}+{B\over r},
\label{19}
\end{equation}
into the RNSF system, where $A$, $B$ and $\gamma$ are constants.\par
The pressure $p\equiv\Pi$ is given by
\begin{equation}
p={{\rho(4ABr^{\gamma+2}-\gamma B^2-\gamma^3\nu^2)}\over{2\gamma r^2}}+
{{\rho A^2 r^{2(\gamma+1)}}\over{2(\gamma+1)}},
\label{20}
\end{equation}
apart from an arbitrary function of time, while the RNSF system yields 
for $U_3$:
\begin{eqnarray}
&&U_3=\exp(-\sigma_0 t)r^{\gamma\over 2}\times\nonumber\\
&&\left[A_0 J_{\gamma\over 2}\left(\sqrt{{\sigma_0}\over\nu}r\right)+
  A_1 Y_{\gamma\over 2}\left(\sqrt{{\sigma_0}\over\nu}r\right)\right],
\label{19a}
\end{eqnarray}
where $\sigma_0>0$, $A_0$, $A_1$ are constants.\par
An interesting physical interpretation of the vortex solution presented in 
{\it{Case ii)}} is the following. Let us consider a device constituted by
two rotating porous coaxial cylinders of radii $r_1<r_2$ containing
in the gap a fluid of kinematic viscosity $\nu$.\par
Let $\gamma={{u_0 r_1}\over\nu}$ be the radial Reynolds number, and 
$u_0\equiv U_1(r_1)$ the radial velocity through the wall of the inner 
cylinder \cite{ML}. We assume that the flow is inward for $\gamma<0$ and 
outward for $\gamma>0$. For $A_0=A_1=0$, the velocity filed reproduces the
solution investigated in \cite{ML}. Furthermore, for $\gamma=0$ the quantities
(\ref{19}) correspond to the Couette flow (\ref{10}). In the case of porous 
cylinders, $A$ and $B$ are related to the geometry and the dynamics of the 
device, and explicitly become: 
$A=-[\Omega_1 a^2(1-b/a^2)]/[r_2^\gamma(1-a^{\gamma+2})]$ 
and $B=r_1^2\Omega_1(1-ba^\gamma)/(1-a^{\gamma+2})$, where $a=r_1/r_2$ and 
$b=\Omega_2/\Omega_1$.\par
In general, i.e. for $U_3\neq 0$, the evolution equation (\ref{r5}) for the 
temperature is very complicated and the finding of exact solutions is a 
difficult task. However, for $A_0=A_1=0$, we obtain
\begin{eqnarray}
&&T=\exp(-\mu_0 t)r^{{\gamma\nu}\over{2k_H}}\times\nonumber\\
&&\left[c_1 J_{{\gamma\nu}\over{2k_H}}\left(\sqrt{{\mu_1}\over{k_H}}r\right)+
  c_2 Y_{{\gamma\nu}\over{2k_H}}\left(\sqrt{{\mu_1}\over{k_H}}r\right)
  \right]\times\nonumber\\
&&\left[c_3\sin\left(\sqrt{{\mu_0-\mu_1}\over{k_H}}z\right)+
  c_4\cos\left(\sqrt{{\mu_0-\mu_1}\over{k_H}}z\right)\right]+T_d^{II},
\label{24}
\end{eqnarray}
where 
\begin{eqnarray}
&&T_d^{II}=-{{2\eta(B^2+\gamma^2\nu^2)}\over{\rho C_p(2k_H+\gamma\nu)r^2}}+
  {{4\eta AB}\over{\gamma\rho C_p(k_H-\nu)}}r^\gamma+\nonumber\\
&&\mu_2 r^{{\gamma\nu}\over{k_H}}-{{\gamma^2\eta A^2 r^{2(\gamma+1)}}\over
  {2\rho C_p(\gamma+1)[2k_H(\gamma+1)-\gamma\nu]}}+T_0^{II}\label{25}
\end{eqnarray}
Here $\mu_0>\mu_1>0$, $\mu_2$ and $T_0^{II}$ are constants, the Prandtl number 
${\rm{Pr}}=\nu/k_H$ is supposed to be $\neq 1$ and $\gamma>0$ (i.e. the 
radial flow is considered outward) to avoid singularities.\par
To determine the time behavior of the coordinates $(x(t),y(t),z(t))$ of the
fluid particle, we integrate the velocity field ${\bf{u}}=(u_1,u_2,u_3)$ to 
get $r^2=2Ct+r_0^2$, where $x$ and $y$ obey the pair of equations of the 
time-dependent oscillator type which afford the solutions 
$x=\sqrt{\tau}\cos\psi$, $y=\sqrt{\tau}\sin\psi$ where $\tau=2Ct+r_0^2$ and 
$\psi={1\over{2C}}[{{2A}\over{\gamma+2}}\tau^{{\gamma\over 2}+1}+
B\ln{\tau\over{\tau_0}}]$.\par
\begin{figure}[htb]
\begin{center}
\epsfig{file=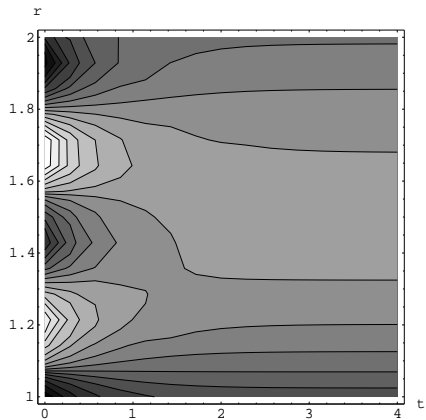,height=5.6cm,width=5.6cm}
\caption{As in Fig. {\ref{fig1}} for the case of porous cylinders 
         ($\gamma=1$).}\label{fig2a}
\end{center}
\end{figure}

The term (\ref{25}) is again due to the presence of the dissipation rate in
(\ref{2}), and represents the asymptotic limit of the solution (\ref{24}), for 
$t\rightarrow\infty$. In Fig. \ref{fig2a} we plot the temperature field 
(\ref{24}) for $z=0.5$. This plot is analogous to that of Fig. (\ref{fig1}). In
other words, also in this case a maximum of $T$ emerges after a transient 
phase.\par
\begin{figure}[htb]
\begin{center}
\epsfig{file=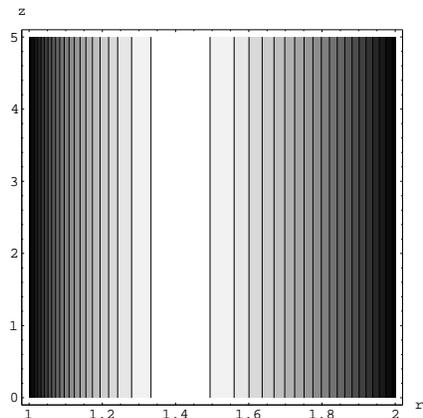,height=5.6cm,width=5.6cm}
\caption{Asymptotic behavior of the temperature field between two rotating
         porous cylinders as a function of $r$ ($1\leq r\leq 2$) and $z$ 
	 ($0\leq z\leq 5$) for $\Delta T=0$, $\gamma=5$ and 
	 $b={{\Omega_2}\over{\Omega_1}}=0.2$ ($Ta=100$).}\label{fig2}
\end{center}
\end{figure}

The position at which $T_{\rm{max}}$ is located in the gap $r_1\leq r\leq r_2$
(Fig. \ref{fig2}) depends on the value of $\gamma$, on the difference of
temperature $\Delta T$ and quadratically on the width of the gap. In particular,
$T_{\rm{max}}$ tends to migrate towards the well of the outer cylinder when 
the values of $\gamma$ increase and vice-versa. A similar situation occurs 
when $\Delta T$ changes.\par
\begin{figure}[htb]
\begin{center}
\epsfig{file=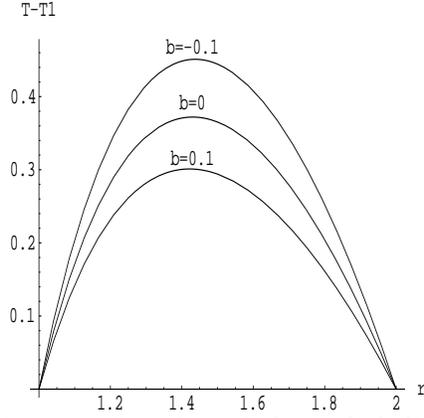,height=5.6cm,width=5.6cm}
\caption{Temperature behavior as a function of the radial distance between the
         two rotating porous cylinders for several values of 
	 $b={{\Omega_2}\over{\Omega_1}}$ and $\gamma=1$ if the cylinders 
	 temperatures are kept equal ($Ta=100$).}\label{fig4}
\end{center}
\end{figure}
\begin{figure}[htb]
\begin{center}
\epsfig{file=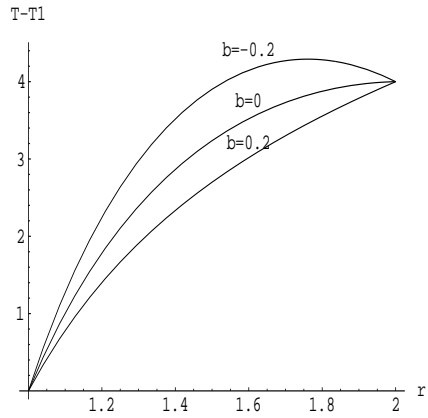,height=5.6cm,width=5.6cm}
\caption{As in Fig. \ref{fig4}, where now $\Delta T=4$ and $\gamma=1$ 
         ($Ta=200$).}\label{fig5}
\end{center}
\end{figure}

In Figs. \ref{fig4} and \ref{fig5} we plot the thermal field for 
$b={{\Omega_2}\over{\Omega_1}}\geq 0$ and $b<0$, when $\Delta T=0$ and 
$\Delta T=4$.
%-------------------------------------------------
%-------------------------------------------------
%-------------------------------------------------
%-------------------------------------------------
%-------------------------------------------------
To conclude, from our analysis of the temperature behavior of vortices of the 
NSF model (\ref{1})-(\ref{2}), notable thermal effects arise as a consequence 
of the presence of the energy dissipation term in the evolution equation
(\ref{2}) for the temperature field. We hope that our theoretical results can
stimulate some experimental work addressed to a possible evidence of such
effects.

\end{document}